\documentclass[twocolumn, superscriptaddress]{revtex4-1}
\usepackage{xfrac, amsmath, graphicx, verbatim, color, hyperref, subfigure}
\usepackage[font=small, labelfont=bf, margin=5mm, format=hang, justification=centerlast]{caption}

\begin{document}

\title{Inflationary Dynamics Reconstruction via Inverse-Scattering Theory}

\author{Jorge Mastache}
\email{mastache@pitt.edu}
\affiliation{CONACYT-Mesoamerican Centre for Theoretical Physics, Universidad Aut\'{o}noma de Chiapas, \\ Carretera Zapata Km. 4, Real del Bosque, 29040, Tuxtla Guti\'{e}rrez, Chiapas, M\'{e}xico.}
\affiliation{Department of Physics and Astronomy, University of Pittsburgh, Pittsburgh, PA 15260 USA and \\ Pittsburgh Particle Physics, Astrophysics, and Cosmology Center (PITT PACC), Pittsburgh PA 15260}

\author{Fernando Zago}
\email{fernando.zago@pitt.edu}
\affiliation{Department of Physics and Astronomy, University of Pittsburgh, Pittsburgh, PA 15260 USA and \\ Pittsburgh Particle Physics, Astrophysics, and Cosmology Center (PITT PACC), Pittsburgh PA 15260}

\author{Arthur Kosowsky}
\email{kosowsky@pitt.edu}
\affiliation{Department of Physics and Astronomy, University of Pittsburgh, Pittsburgh, PA 15260 USA and \\ Pittsburgh Particle Physics, Astrophysics, and Cosmology Center (PITT PACC), Pittsburgh PA 15260}

\begin{abstract}
The evolution of inflationary fluctuations can be recast as an inverse scattering problem. In this context, we employ the Gel'fand-Levitan method from inverse-scattering theory to reconstruct the evolution of both the inflaton field freeze-out horizon and the Hubble parameter during inflation. We demonstrate this reconstruction procedure numerically for a scenario of slow-roll inflation, as well as for a scenario which temporarily departs from slow-roll. The field freeze-out horizon is reconstructed from the accessible primordial scalar power spectrum alone, while the reconstruction of the Hubble parameter requires additional information from the tensor power spectrum. We briefly discuss the application of this technique to more realistic cases incorporating estimates of the primordial power spectra over limited ranges of scales and with specified uncertainties.
\end{abstract}


\maketitle

\section{Introduction}\label{sec:Introduction}

An early period of accelerated cosmic expansion, referred to as inflation, can successfully account for the observed large-scale properties of our Universe  \cite{Guth, Sato}. The inflationary paradigm also provides a natural and appealing mechanism for the generation of density perturbations with the correct statistical properties to seed the growth of structure in the Universe via gravitational instability (see Refs.~\cite{Linde, Planck_Inflation} for a review). However, despite the concordance between observations and the predictions of an inflationary scenario, the physics responsible for driving inflation remains largely speculative.

The reconstruction program for inflation has the goal of probing the physics behind the inflationary expansion by constraining the process of inflation from observable features of the Universe. The most direct source of information about the inflationary epoch is the primordial power spectrum of density perturbations, $P_{\mathcal{R}}(k)$, and, if eventually detected, of tensor perturbations, $P_T(k)$. The importance of the primordial density perturbations is due to their origin as quantum fluctuations in the ``inflaton" field driving inflation, while it is generally thought that quantum fluctuations in the gravitational field during inflation will produce tensor perturbations. The expansion history during inflation is imprinted in both. Measurements of the primordial power spectra over a sufficiently wide range of scales will strongly constrain the possible inflation expansion histories \cite{Caligiuri:2015} and thus physics at energy scales far higher than are accessible in any conceivable accelerator experiment. Current measurements of the microwave background temperature anisotropies show that the primordial density power spectrum is a power law over an order of magnitude in wave number \cite{Hlozek,Planck_Inflation}, and the amplitude and slope of the power spectrum already rules out a range of inflation models \cite{Planck_Inflation}.

Drawing conclusions about inflation dynamics from measurements (or imagined future measurements) has typically been done in one of two ways. The first is with a perturbative expansion around slow-roll inflaton dynamics, which gives constraints on the slow-roll parameters given the amplitude and power-law index of  scalar and tensor perturbations at the scale of the horizon today \cite{Planck_Inflation}. A measurement of the tensor amplitude at solar-system scales would provide a much longer lever arm and additional slow-roll parameter constraints \cite{Caligiuri}. The fact that both scalar and tensor perturbations arise from the same inflation dynamics gives consistency relations between the amplitudes and power laws which must be satisfied if inflation dynamics are in the slow-roll regime \cite{Copeland:1998fz,Steinhardt:1984jj}. While confirmation of these consistency relations would constitute an impressive success of the inflation idea, this type of analysis can reveal inflation dynamics across a relatively limited portion of the inflationary epoch since it is essentially an expansion around the time when perturbations on the scale of the present horizon were generated. Moreover, extending the slow-roll expansion to second order leads to non-perturbative corrections in the reconstructed inflation dynamics \cite{deOliveira2005mf}, indicating lack of convergence in the slow-roll expansion. A second more general technique is to generate inflation models with a large range of expansion histories, then pick out the ones which are consistent with a given set of perturbation spectra measurements \cite{Caligiuri:2015, Kinney:2002qn}. This technique can incorporate entire inflation histories, but it is difficult to quantify the resulting constraints on inflation because there is no natural prior probability on the space of inflation models \cite{Kinney:2002qn}.

Instead of picking inflation models and seeing if they satisfy some set of measurements, an alternative possibility is to do the reverse: find the constraints on the inflation history which arise directly from the measurements. While calculating power spectra given an inflation model is straightforward, inferring an inflation model given a density power spectrum is more challenging. Habib et al.~\cite{Habib} pointed out that this problem is formally analogous to inverse-scattering theory in quantum mechanics. 

The task of inverse-scattering theory is to determine the features of the scattering target, given the distant-past input and far-future scattered output waves. In the context of inflation, the early time evolution of the incoming wave functions are set by the Bunch-Davies vacuum conditions and the late time scattered wave functions are characterized by a freeze-out behavior as they become super-horizon. This formalism identifies the inflaton freeze-out horizon as the effective scattering potential and the primordial power spectrum as the scattering data. The freeze-out horizon can then be used to reconstruct the evolution of the scale factor and Hubble parameter during inflation. This last step requires the amplitude of the tensor power spectrum at a fiducial scale.

A formal solution to the quantum inverse scattering problem was developed by Gel'fand, Levitan, and Marchenko \cite{Gelfand_Levitan, Marchenko, Marchenko11}. In this paper, we implement a numerical solution to the Gel'fand-Levitan-Marchenko equation in the context of inflation, given exact knowledge of the primordial density power spectrum. We demonstrate the recovery of both a purely slow-roll inflation model and an inflation model with a brief period of fast-roll evolution. This calculation provides the basic groundwork for inferring an inflation model based on partial data about the primordial density power spectrum, possibly combined with measurement of the primordial tensor power spectrum on one or more scales. It has the advantage of relying solely on measurable quantities, making no assumption about the form of the inflaton potential and thus providing model-independent constraints on inflation assuming only the standard connection between perturbation amplitude and scale factor evolution.

In Section \ref{sec:Inflationary_Fluctuations} we briefly review the relevant expressions for the evolution of primordial fluctuations in a single scalar field inflationary scenario. In Section \ref{sec:Inflation_Inverse_Scattering} the evolution of inflationary fluctuations is recast as an inverse-scattering problem and the Gel'fand-Levitan method for the inversion of scattering problems is introduced.  Section \ref{sec:Results} is devoted to a numerical solution of the inverse-scattering problem in the context of inflation, resulting in the reconstruction of the field freeze-out horizon and the Hubble parameter. Finally, in the concluding section we discuss the application of these numerical techniques to more realistic cases incorporating estimates of the primordial power spectra over limited ranges of scales and with specified uncertainties. Some calculation details are summarized in the Appendix. Natural units with $\hbar = c = 1$ are adopted throughout.

\section{Review of Inflationary Fluctuations}\label{sec:Inflationary_Fluctuations}

Here we briefly review the relevant results in the analysis of primordial inflationary fluctuations in a single-field inflationary scenario. In solving the perturbed Einstein equations, the gauge invariant variable $\phi$ arises which incorporates fluctuations of both the metric and inflaton field. This quantity can be decomposed in its Fourier modes, $\phi_k$, which evolve according to the Mukhanov-Sasaki equation \cite{Mukhanov:1990me}:
\begin{equation}\label{eq:Mode_Evolution_Equation}
	\phi^{\prime\prime}_{k} + [k^{2} - q(\eta)]\phi_{k} = 0 \, ,
\end{equation}
where the primes indicate differentiation with respect to conformal time $\eta \equiv \int a^{-1}dt $, $k$ is the comoving wavenumber, and $q(\eta)$ corresponds to the field freeze-out horizon. For scalar modes $q(\eta) = z^{\prime\prime}/z$, where $z$ is the Mukhanov variable defined by
\begin{equation}\label{eq:Mukhanov_Variable}
	z \equiv \frac{a(\eta)\sqrt{2\epsilon(\eta)}}{c_{s}} \, ,
\end{equation}
with $\epsilon \equiv -H^{\prime}/aH^{2}$ the the slow-roll parameter, $H$ the Hubble parameter, and $c_s$ the adiabatic sound speed during inflation. For simplicity we assume $c_s = 1$, appropriate for inflation driven by a scalar field.

The evolution of each mode goes through two main phases, sub-horizon and super-horizon evolution. During sub-horizon evolution, the mode wavelength is small compared to the horizon size, $k \gg q(\eta)$. From Eq.~(\ref{eq:Mode_Evolution_Equation}) this condition implies that each mode oscillates as in a non-expanding Universe. This allows the initial conditions for each mode $\phi_{k}$ to be set through an adiabatic correspondence with the positive frequency modes of Minkowski vacuum, thus fixing $\phi_k \sim (2k)^{-\frac{1}{2}}e^{-ik\eta}$ for $\eta \rightarrow -\infty$. This is the well-known Bunch-Davies vacuum initial condition;  each sub-horizon mode evolves to a good approximation as a free mode in flat space.

However, as inflation unfolds $q(\eta)$ increases, and subhorizon modes eventually become super-horizon with $k \ll q(\eta)$. In this limit the mode $\phi_k$ no longer propagates freely, but rather exhibits a freeze-out behavior given by $\phi_k \sim A_k z(\eta)$, where $A_k$ is a function of $k$ alone. It is also in this limit that the gauge-invariant curvature fluctuations, $\mathcal{R}_{k} = -\phi_{k}/z$, freeze out at the constant value $\mathcal{R}_{k} = -A_{k}$.

The quantity $A_k$ is directly related to the primordial power spectrum of scalar fluctuations
\begin{equation}\label{eq:Primordial_PS}
	P(k) = \frac{k^3}{2\pi^2}|A_k|^2 \, .
\end{equation}
The value of a perturbation, given by $A_k$ for a wavenumber $k$, is to a good approximation fixed at horizon crossing $k\simeq q(\eta)$. Consequently, a measurement of $P(k)$ constrains the evolution of the horizon crossing scale  $q(\eta)$ during inflation. The power spectrum defined above is usually expressed as a dimensionless quantity $P_{\mathcal{R}}(k) = 2\pi G \, P(k)$, where $G$ is the gravitational constant. In the next Section, we demonstrate how $q(\eta)$ can be recovered from $P_{\mathcal{R}}(k)$.

In addition to scalar perturbations, inflation generates tensor perturbations whose evolution can also be understood in terms of the freeze-out formalism discussed above. Just like the scalar modes, the tensor modes $h_k$ evolve according to Eq.\,(\ref{eq:Mode_Evolution_Equation}), with their freeze-out horizon assuming a simplified form given by $q(\eta) = a^{\prime\prime}/a$. A power spectrum of primordial tensor perturbations is defined in a similar fashion from the freeze-out of tensor modes given by $h_k \sim B_k a(\eta)$, and by accounting for their two polarization states:
\begin{equation}\label{eq:Primordial_PT}
	P_T(k) = 2\frac{k^3}{2\pi^2}|B_k|^2 \, .
\end{equation}
 In particular, in the slow-roll limit ($\epsilon$ nearly constant),  the freeze-out horizon is $q(\eta) \simeq 2\eta^{-2}$ and an exact solution can be computed for the evolution of each mode. In this regime the tensor power spectrum assumes the form $P_T(k) \left. \propto H^2 \right |_{k=aH}$.  A measurement of $P_T(k)$ at a particular scale further allows determination of the expansion history $a(\eta)$ during inflation, given the freezeout horizon $q(\eta)$. 

\section{Inflation as an Inverse-Scattering Problem}\label{sec:Inflation_Inverse_Scattering}

Following Ref.~\cite{Habib}, we rephrase the evolution of inflationary perturbations as a scattering problem and discuss how the Gel'fand-Levitan method from inverse-scattering theory can be used to reconstruct inflation dynamics. First, make the two following substitutions in Eq.~(\ref{eq:Mode_Evolution_Equation}):
\begin{equation}\label{eq:Scattering_Substitutions}
r = -\eta \ \ \  \mathrm{and}  \ \ \ q(-r) = V(r) + \ell(\ell+1)r^{-2} \, .
\end{equation}
The result resembles a radial time-independent Schr\"{o}dinger equation for a wave-function $\phi(k, r)$ interacting with a central potential $V(r)$:

\begin{equation}\label{eq:Schrodinger_Scattering_Equation}
	\phi^{\prime\prime}(k,r) + \left[k^2 - \frac{\ell(\ell+1)}{r^2} - V(r) \right]\phi(k,r) = 0 \, ,
\end{equation}
where $\ell$ is a constant determined by the form of the primordial power spectrum $P_{\mathcal{R}}(k)$.

From the Bunch-Davies vacuum choice, it also follows that each wave-function behaves as an incoming wave $\phi(k,r) \sim e^{ikr}$ for $r \rightarrow \infty$. In this picture, every mode $\phi_k$ is now represented by an incident particle of energy $k^2$ and angular momentum $\ell$ scattering off of a central potential. The description of inflationary fluctuations as the evolution of modes of different scales has therefore been replaced by the scattering of fictitious particles with varying energies and fixed angular momentum. The mathematical treatment for this class of scattering problem is well known from inverse-scattering theory.

To apply inverse-scattering techniques, the potential must satisfy the regularity condition
\begin{equation}\label{eq:Regular_Condition_Potential}
	\int_{b}^{\infty}| V(r) | r dr < \infty \ \ \ \mathrm{for} \ \ \ b \geq 0 \, .
\end{equation}
This expression imposes that $V(r)$ must decrease faster than $r^{-2}$ as $r \rightarrow \infty$. This will always hold for single field inflation as long as the expansion is nearly exponential in the deep past. Indeed, in this case it is known that the freeze-out horizon behaves approximately as $q \simeq 2r^{-2}$ for large values of $r$. From (\ref{eq:Scattering_Substitutions}) it then follows that the ``centrifugal" term $\ell(\ell + 1)r^{-2}$ retains the relevant functional form in the large $r$ regime, implying that the potential must fall faster than $r^{-2}$ when $r \rightarrow \infty$. Eq.~(\ref{eq:Regular_Condition_Potential}) also imposes a degree of regularity on the behavior of $V(r)$ for $r \rightarrow 0$, but the form of the potential near the origin is not observationally accessible, as it can only be probed by particles (or modes) with extremely large values of $k$. Thus, $V(r)$ can be assumed to be regular for $r \rightarrow 0$ without significantly affecting the reconstruction procedure. This will become particularly evident once the Gel'fand-Levitan formalism is stated in the discussion that follows. This shows that a potential $V(r)$ associated with a general single field inflation model can be regarded as regular over the entire range of interest and, consequently, that inverse-scattering tools apply without restrictions in the context of single field inflation.

The task of inverse-scattering theory is to determine the shape of a scattering potential, $V(r)$, from the information contained in the scattered wave-functions. This information is encapsulated in the so-called Jost function, denoted by $F_{\ell}(k)$, which is constructed from a specific combination of ingoing and outgoing scattering solutions. Habib et.~al \cite{Habib} showed that, in the context of inflation, the Jost function could be related to $P_{\mathcal{R}}(k)$ through the following expression (see Appendix \ref{app:Jost_Function} for more details):
\begin{equation}\label{eq:Jost_Function}
	\left| F_{\ell}(k) \right|^2 = \frac{r_0^{2}\pi^{2}}{8 \, G \left| \Gamma\left( \ell + \frac{1}{2} \right) \right|^{2} } \left(  \frac{k}{2}  \right)^{2\ell - 2} P_{\mathcal{R}}(k) \, ,
\end{equation}
where both $r_0$ and $\ell$ are constants of the scattering problem determined by the primordial power spectrum $P_{\mathcal{R}}(k)$. The Jost function also satisfies the asymptotic condition $\lim_{k \to \infty}\left|F_{\ell}(k)\right| = 1$ (see Refs. \cite{Newton, Newton_Chadan_Sabatier, Moser_Baltes, Ablowitz_Segur} for detailed treatments).

This method provides a solution to the Gel'fand-Levitan-Marchenko equation
\begin{equation}\label{eq:GLM_Equation}
	K(r,s) + G(r,s) + \int_0^r K(r,t) G(t,s) dt = 0 \, ,
\end{equation}
which relates the input kernel
\begin{equation}\label{eq:Input_Kernel}
	G(r,s) = \frac{2}{\pi}\int_0^\infty S_{\ell}(r k)S_{\ell}(s k) \left[ |F_{\ell}(k)|^{-2}-1 \right]dk \,
\end{equation}
to the output kernel $K(r,s)$.
Here $\mathit{S}_\ell(x)$ are the Ricatti-Bessel functions defined in terms of the usual spherical Bessel functions as $\mathit{S}_\ell(x) = x \mathit{j}_\ell(x)$. Equation (\ref{eq:GLM_Equation}) is a Fredholm integral equation of the second kind which can be solved for $K(r,s)$ by a separable-kernel decomposition, as explained in Appendix \ref{app:Solving_GLM}. Once $K(r,s)$ is obtained, the scattering potential is given by
\begin{equation}\label{eq:Reconstructed_Potential}
	V(r) = 2\frac{d}{dr}K(r,r) \, .
\end{equation}
The scattering  potential determines directly the inflation freeze-out horizon $q(\eta)$ by Eq.~(\ref{eq:Scattering_Substitutions}). 

Once $q(\eta)$ is determined, the Mukhanov variable is the solution to
\begin{equation}
\label{eq:FreezeOut_Diff_Equation}
	z^{\prime\prime} = q(\eta)z
\end{equation}
and the scale factor satisfies
\begin{equation}
\label{eq:Scale_Factor_Diff_Equation}
	a^{\prime\prime} = 2\frac{{a^{\prime}}^2}{a} - \frac{z(\eta)^2}{2}\frac{{a^{\prime}}^2}{a^3} \, , 
\end{equation}
which follow from the definitions of $q(\eta)$ and $z(\eta)$. It is convenient to change the independent variable in these differential equations to $\beta = \ln(-\eta)$ in order to facilitate their numerical treatment.

The solutions for both (\ref{eq:FreezeOut_Diff_Equation}) and (\ref{eq:Scale_Factor_Diff_Equation}) require initial conditions for $z(\eta)$ and $a(\eta)$. It can be shown, however, that the two initial conditions follow from the amplitude of the tensor power spectrum at CMB scales, where the inflaton field is expected to be slow-rolling and the primordial tensor power spectrum has the form $P_T \propto H^{2}$. Once these initial conditions are determined, the above equations can be integrated as an initial value problem.

To obtain the initial conditions for $z(\eta)$ and $a(\eta)$ from $P_T(k)$ and the slow-roll conditions, note that at CMB scales the freeze-out horizon is $q = (2 + 3\epsilon)/r^2$ to first order in slow-roll parameters. Since the potential $V(r)$ represents deviations from exactly exponential inflation it must be small at CMB scales, and the freeze-out horizon is dominated by the centrifugal term in Eq.~(\ref{eq:Scattering_Substitutions}), $q = \ell (\ell + 1)/r^2$. Therefore, to lowest order in slow-roll $\ell \approx 1 + \epsilon$ or, equivalently, $\ell \approx (3 - n_s)/2$ in terms of the spectral index, $n_s$. Furthermore, the Mukhanov variable at such scales is $z = r_0/r^\ell$, where $r_0$ is obtained from the power spectrum normalization at a fiducial CMB scale, $k_0$. This can be done by evolving the mode of wavenumber $k_0$ from the distant past through freeze-out and computing $P_{\mathcal{R}}(k_0)$ for different values of $r_0$. Matching this value with the known power spectrum normalization at $k_0$ singles out a value for $r_0$. Since $P_{\mathcal{R}}(k)$ is known at CMB scales, the constants $\ell$ and $r_0$ can be determined from the power spectrum slope and normalization, respectively. Alternatively, one could also determine the value of $r_0$ from the small $k$ expansion of the Jost function derived in \cite{Habib}. Therefore, the initial conditions $z$ and $z^{\prime}$ for Eq.~(\ref{eq:FreezeOut_Diff_Equation}) can be computed from $z = r_0/r^\ell$, where the initial value of $r$ is given by the horizon crossing relation $k(r) = \sqrt{q(r)}$. The initial conditions $a$ and $a^{\prime}$ for Eq.~(\ref{eq:Scale_Factor_Diff_Equation}) then follow from the slow-roll condition $z^{\prime} \simeq (a^{\prime}/a)z$, the definition of the Hubble parameter $H = a^{\prime}/a^{2}$, and a measurement of $P_T \propto H^{2}$ at CMB scales.

Thus, the necessary initial condition for reconstructing the Mukhanov variable and the scale factor during inflation, and by consequence the Hubble parameter, can be obtained from a measurement of the amplitude of the primordial tensor power spectrum at CMB scales.



\begin{figure}
    \includegraphics[width=0.45\textwidth]{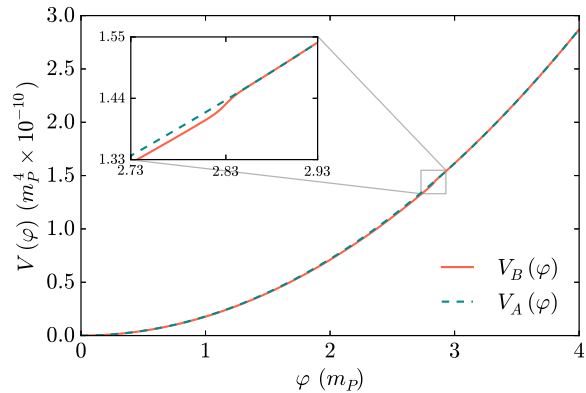}
    \caption{Plot of the inflaton potentials used to obtain the simulated primordial power spectra. The dashed line shows the quadratic potential, $V_A(\varphi)$, while the solid line corresponds to the potential $V_B(\varphi)$, in Eq.~(\ref{eq:Quadratic_Potential_Feature}).}
    \label{fig:inflaton_potential}
\end{figure}

\begin{figure*}
  \subfigure[\,Quadratic potential]{
    \label{fig:Power_Spectrum_Quadratic}
    \includegraphics[scale=0.45]{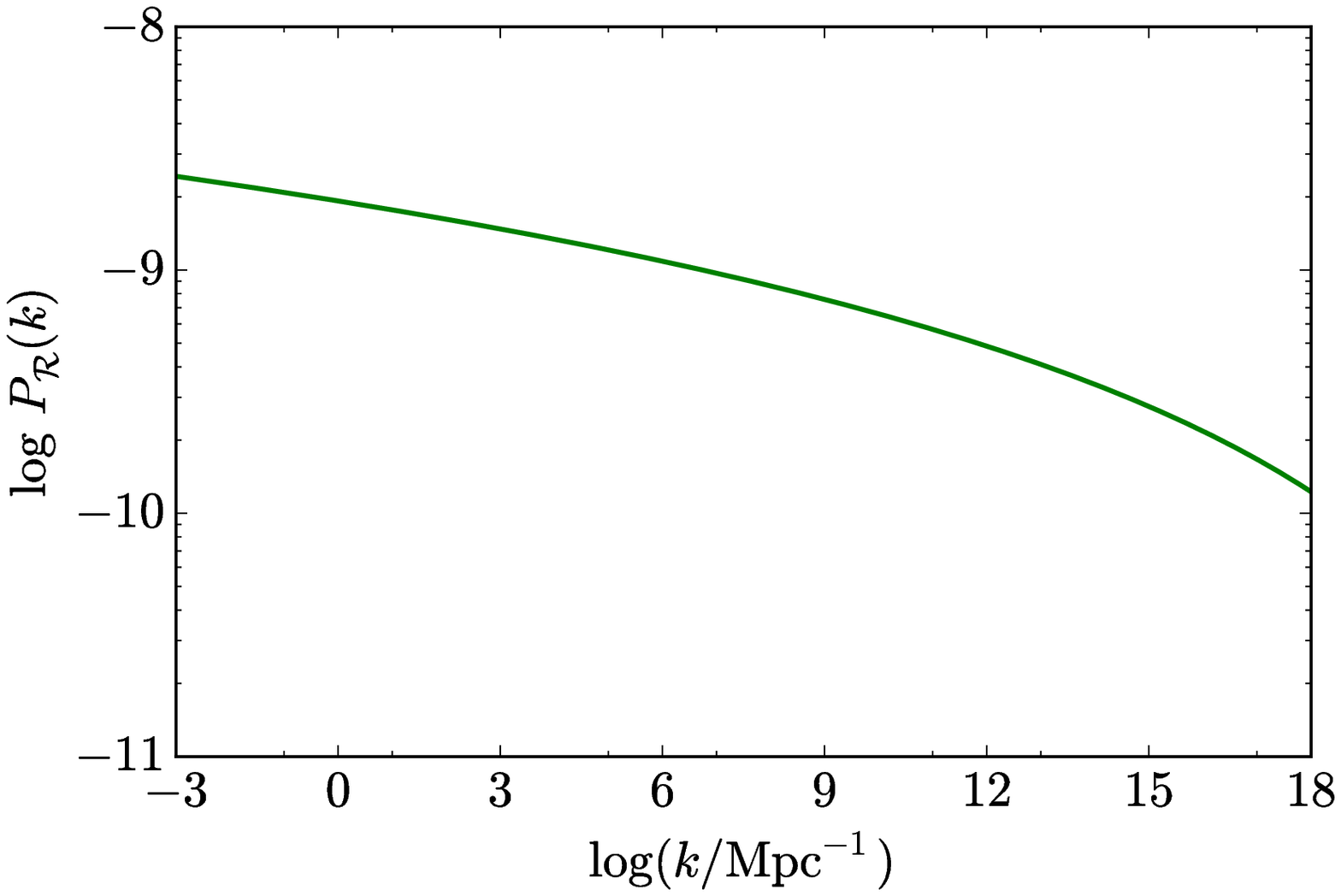}
  }
  \hspace*{\fill}
  \subfigure[\,Potential with a step]{
    \label{fig:Power_Spectrum_Step}
    \includegraphics[scale=0.45]{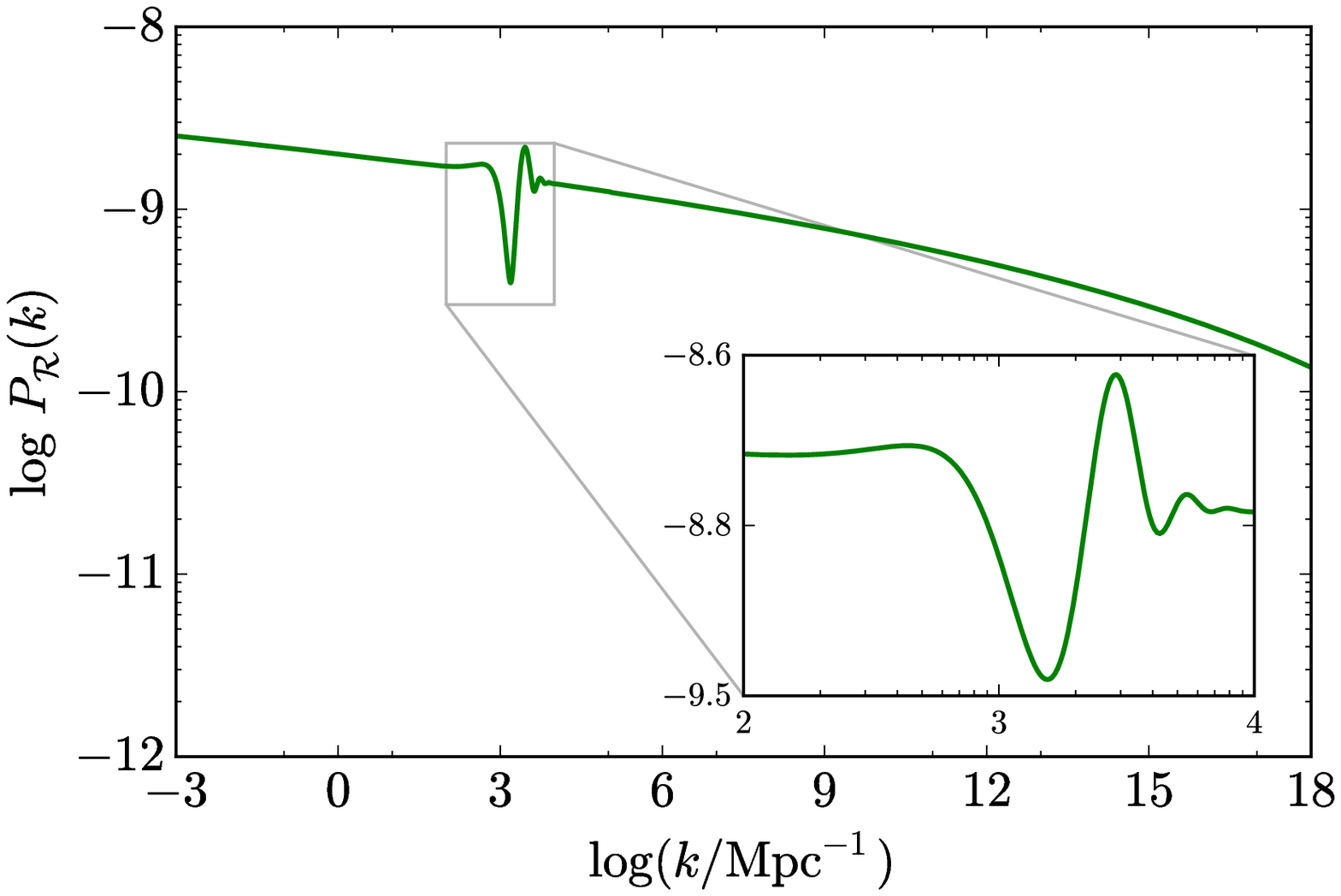}
  }
  \caption{Primordial power spectra associated with the potential $V_{A}(\varphi)$ (left panel) and $V_{B}(\varphi)$ (right panel). The inset focuses on the ringing feature generated by the potential step. Both curves are normalized according to Planck results at the pivot scale $k_0 = 0.05\, \mathrm{Mpc^{-1}}$, such that $P_{\mathcal{R}}(k_0) = 2.196\times 10^{-9}$. 
  }
  \label{fig:Power_Spectrum}
\end{figure*}

\section{Analysis and Results}\label{sec:Results}

Numerical solutions to Eq.~(\ref{eq:GLM_Equation}) provide information about arbitrary inflation dynamics; an assumption of slow-roll behavior only comes in to initial conditions at CMB scales, where observations of $P_{\mathcal{R}}(k)$ show this is likely a good assumption \cite{Planck_Inflation}. We demonstrate numerical solutions in two particular cases, corresponding to single-field inflation models with the scalar potentials
\begin{eqnarray}
	V_{A}(\varphi) &=& m^{2}\varphi^{2} \, ,  \label{eq:Quadratic_Potential}\\
	V_{B}(\varphi) &=& m^{2}\varphi^{2} \left[ 1 + c \tanh\left( \frac{\varphi - \varphi_s}{d} \right) \right] \, . \label{eq:Quadratic_Potential_Feature}
\end{eqnarray}
While the first potential has the standard quadratic form and corresponds to a model in which the inflaton field is slow-rolling for most of its evolution, the second potential possesses the interesting characteristic of allowing the inflaton field to temporarily deviate from slow-roll when $\varphi \approx \varphi_s$. The constants $c$ and $d$ correspond to the amplitude and breadth of a small step in the potential, which is responsible for the departure from slow roll. These potentials are shown in Fig.~\ref{fig:inflaton_potential} and their resulting scalar power spectra $P_{\mathcal{R}}(k)$ in Fig.~\ref{fig:Power_Spectrum}. The primordial power spectra were numerically generated according to the standard mode freeze-out formalism reviewed in Section \ref{sec:Inflationary_Fluctuations} (see \cite{Adams} for a detailed description of the numerical aspects of this calculation).

For a rigorous application of the Gel'fand-Levitan method we should have access to $P_{\mathcal{R}}(k)$ for all $k$. However, in practice our knowledge is limited by the range of scales accessible to our experiments. 
Here we study three scenarios in which we assume access to $P_{\mathcal{R}}(k)$ up to $k_{\star} = 10^{4}$, $10^{6}$, and $10^{8}\, \mathrm{Mpc^{-1}}$. Our choices of cut-off scales span the range over which future information on the primordial power spectrum may be available from various sources including strong lensing \cite{Keeton:2009}, dynamics of tidal streams \cite{Ngan:2014}, and microwave background spectral distortions \cite{Chluba:2012we}. Furthermore, from the asymptotic behavior of the Jost function we notice that any information contained in the spectrum of extremely small scales should not contribute significantly to the reconstruction procedure. This is clearly seen from the integrand of Eq.~(\ref{eq:Input_Kernel}) which tends to zero for large values of $k$.

A simple toy model which illustrates the procedure discussed in Section \ref{sec:Inflation_Inverse_Scattering} is the trivial case of inflation driven by a constant inflaton potential resulting in eternal exponential expansion. In this case the primordial power spectrum is exactly scale invariant $P_{\mathcal{R}}(k) \propto k^{0}$, and the slow-roll parameter vanishes $\epsilon = 0$. Therefore, following our previous discussion, $\ell = 1$. As a result the Jost function, Eq.~(\ref{eq:Jost_Function}), will also be scale invariant $\left| F_{\ell}(k) \right|^2 \propto r_0^2$, where a suitable value for $r_{0}$ can always be found. A constant Jost function implies that both input and output kernels are identically zero, $G(r,s) = K(r,s) = 0$. As a consequence, the scattering potential given by Eq.~(\ref{eq:Reconstructed_Potential}) is $V(r) = 0$. Finally, from Eq.~(\ref{eq:Scattering_Substitutions}) the frezee-out horizon is found to be $q = 2r^{-2}$ which is in agreement with the theoretical result for de Sitter space.

The procedure is analogous for the more complicated inflationary potentials of Eqs.~(\ref{eq:Quadratic_Potential}) and (\ref{eq:Quadratic_Potential_Feature}). The values of $\ell$ and $r_0$ are determined from the scalar power spectrum slope and normalization, respectively. For the cases studied in this work we obtained $\ell = 1.008$ since $\epsilon = 0.008$ when CMB scales exit the horizon in our simulations. The value of $r_0$ follows from the normalization at the pivot scale $k_0 = 0.05\, \mathrm{Mpc^{-1}}$, such that $P_{\mathcal{R}}(k_0) = 2.196\times 10^{-9}$.

The numerical value of the input kernel is found by evaluating the highly oscillatory integral in Eq.~(\ref{eq:Input_Kernel}) up to $k_{\star}$. In this work this computation was performed using the Levin collocation method of integration \cite{Levin} (see Appendix \ref{app:Levin_Method}) which takes into account the fact that the Jost function is, in the most general scenario, a non-smooth function of $k$. The parameters $r$ and $s$ both range over the interval $(0,\,k_\star)$. The expression for the input kernel is closely related to the two dimensional Fourier transform \citep{Marchenko11} of the scattering function in radial coordinates, which is commonly used in inverse-scattering theory.

\begin{figure*}
  \subfigure[\,Quadratic potential]{
    \label{fig:Reconstructed_changing_freezeout_Quadratic}
    \includegraphics[scale=0.45]{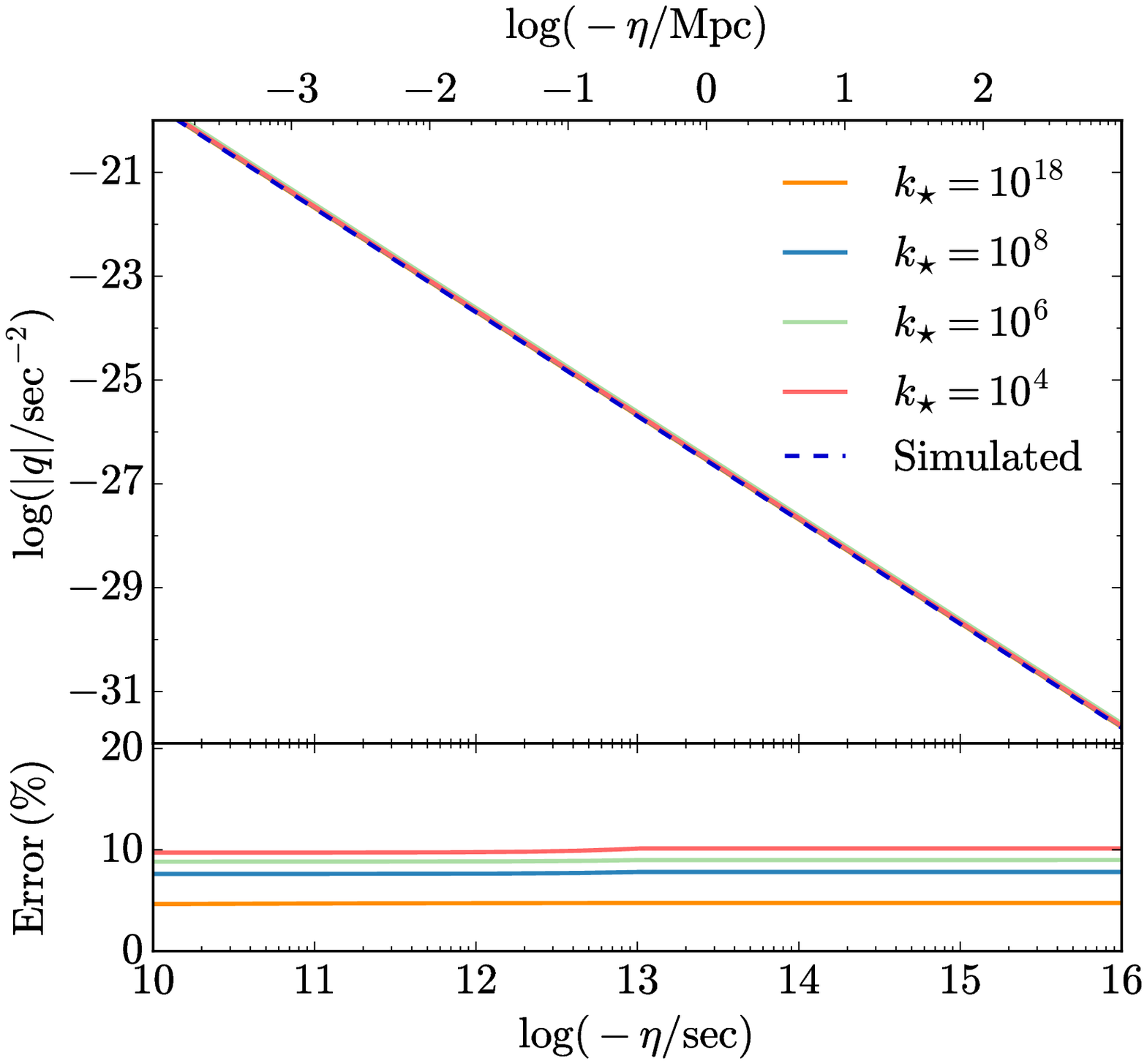}
  }
  \hspace*{\fill}
  \subfigure[\,Potential with a step]{
    \label{fig:Reconstructed_changing_freezeout_Step}
    \includegraphics[scale=0.45]{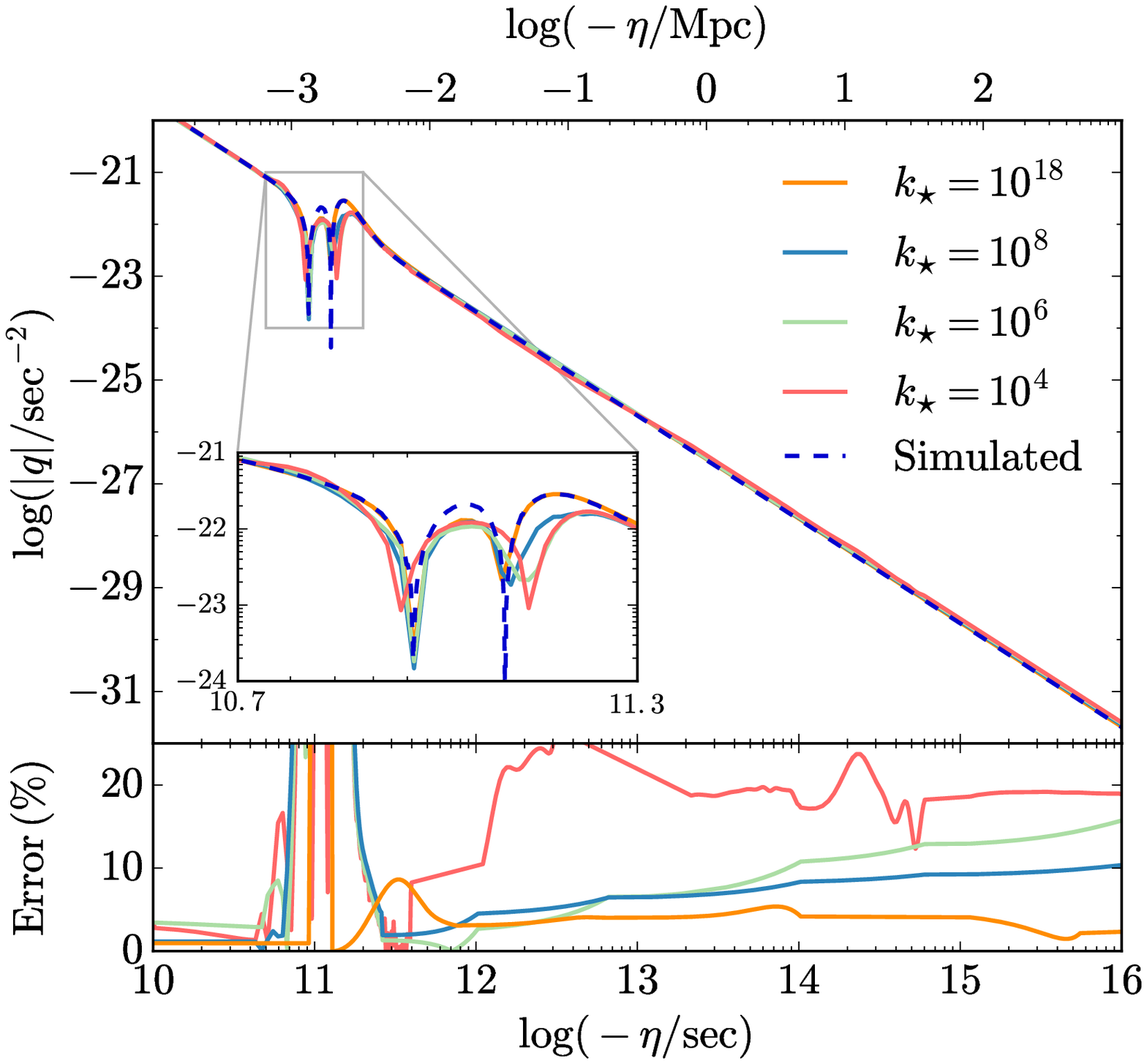}
  }
  \caption{Reconstruction of the freeze-out horizon of $V_{A}(\varphi)$ (left panels) and  $V_{B}(\varphi)$ (right panels) for different choices of the cut-off scale, $k_\star$ (in units of $\mathrm{Mpc^{-1}}$). For the regions where the primordial power spectrum is featureless the reconstruction of $q(\eta)$ resembles the simulated curve.  However, whenever features are present, the corresponding features in the reconstructed $q(\eta)$ are suppressed. The larger the value of $k_\star$ the better the reconstruction, as can be seen for the case with the inflaton potential $V_{B}(\varphi)$. The case with cut-off $k_\star = 10^{18} \, \mathrm{Mpc^{-1}}$, although not physically well-motivated, is shown to demonstrate the convergence of the reconstruction procedure. The inset on the right panel shows a close-up of the feature. The difference between the simulated and the reconstructed curves is shown in both lower panels.
  }
  \label{fig:Reconstructed_changing_freezeout}
\end{figure*}

The reconstruction procedure is then carried out by employing the inverse-scattering formalism. The output kernel follows as a solution of the Gel'fand-Levitan-Marchenko equation, which can be rewritten as a set of linear equations assuming the input kernel is separable,
 \begin{equation}
	 G(r,s) = \sum_{i} a_{i}(r) b_{i}(s) \, .
 \end{equation}
A detailed derivation is presented in Appendix \ref{app:Solving_GLM}. The reconstructed freeze-out horizon obtained from this analysis for both model potentials in Eqs.~(\ref{eq:Quadratic_Potential}) and (\ref{eq:Quadratic_Potential_Feature}) are shown in Fig.~\ref{fig:Reconstructed_changing_freezeout} for cut-off scales $k_{\star} = 10^{4}$, $10^{6}$, and $10^{8}\, \mathrm{Mpc^{-1}}$. Along with these physically well-motivated choices of $k_{\star}$ we also plot the reconstructed curve obtained with $k_\star = 10^{18} \, \mathrm{Mpc^{-1}}$ as a mere demonstration of numerical convergence of the reconstruction procedure with increasing cut-off scales.

Figure \ref{fig:Reconstructed_changing_freezeout_Quadratic} shows that in the case of a featureless power spectrum, choosing a cut-off scale of $k_{\star} = 10^{4} \, \mathrm{Mpc^{-1}}$ preserves the overall functional form of the reconstruction, but induces an average relative error which is approximately 2\,\% larger compared to that obtained for $k_{\star} = 10^{8} \, \mathrm{Mpc^{-1}}$. We obtained average errors of 9.7\,\%, 8.8\,\%, and 7.6\,\% for $k_{\star} = 10^{4}$, $10^{6}$, and $10^{8}\, \mathrm{Mpc^{-1}}$, respectively.

On the other hand, if the primordial power spectrum possesses features such as the ringing oscillations shown in Fig.~{\ref{fig:Power_Spectrum_Step}}, the corresponding feature in the reconstructed freeze-out horizon is suppressed and broadened for small cut-off scales. This can be seen in Fig.~\ref{fig:Reconstructed_changing_freezeout_Quadratic}. The quality of our results for this case depends on whether the reconstructed region contains the feature. The average error in the region which does not contain the feature is of 6.7\,\% for our largest choice of cut-off scale $k_{\star} = 10^{8} \, \mathrm{Mpc^{-1}}$, while for $k_{\star} = 10^{6}$ and $10^{8}\, \mathrm{Mpc^{-1}}$ the average errors are of 7.2\,\%, and 11.4\,\%, respectively. These errors, although dependent on the choice of cut-off, are expected to decrease with higher numerical resolution. The region containing the feature is harder to reconstruct and represents a difficult numerical challenge. In this region, the average percentage difference between the simulated and reconstructed curves for $k_\star = 10^{8} \,\mathrm{Mpc^{-1}}$ is of 48\,\%. Bringing the cut-off scale down to $k_{\star} = 10^{6}$ and $10^{4}\, \mathrm{Mpc^{-1}}$ increases the average relative error to 72\% and 75\%, respectively. However, the characteristic shape of the feature is reproduced successfully.

The inverse-scattering formalism shows that we can recover the freeze-out horizon, $q(\eta)$, from the scalar power spectrum alone. To reconstruct the expansion history, however, additional information is needed for the integration of Eqs.~(\ref{eq:FreezeOut_Diff_Equation}) and (\ref{eq:Scale_Factor_Diff_Equation}). In principle, as explained in Section \ref{sec:Inflation_Inverse_Scattering}, the amplitude of $P_T(k)$ at CMB scales could provide the necessary initial conditions for the numerical solution of these equations, allowing us to reconstruct $a(\eta)$ and $z(\eta)$. We have reconstructed $H$ from the freeze-out horizon assuming that the amplitude of the primordial tensor power spectrum is known at $k_0 = 0.05\, \mathrm{Mpc^{-1}}$. The numerical errors in the reconstructed freeze-out horizon make the initial step in the integration of Eq.~(\ref{eq:FreezeOut_Diff_Equation}) inconsistent with the initial conditions derived for the Mukhanov variable at CMB scales. In order to eliminate this inconsistency we rescale the reconstructed freeze-out horizon by a constant factor to ensure it agrees with its known slow-roll value at CMB scales, $q = \ell(\ell + 1)\eta^{-2}$. This operation preserves the overall shape of the freeze-out horizon and also guarantees consistency in the integration of Eq.~(\ref{eq:Scale_Factor_Diff_Equation}).

\begin{figure*}
  \subfigure[\,Quadratic potential]{
    \label{fig:Reconstructed_Hubble_Parameter_Quadratic}
    \includegraphics[scale=0.45]{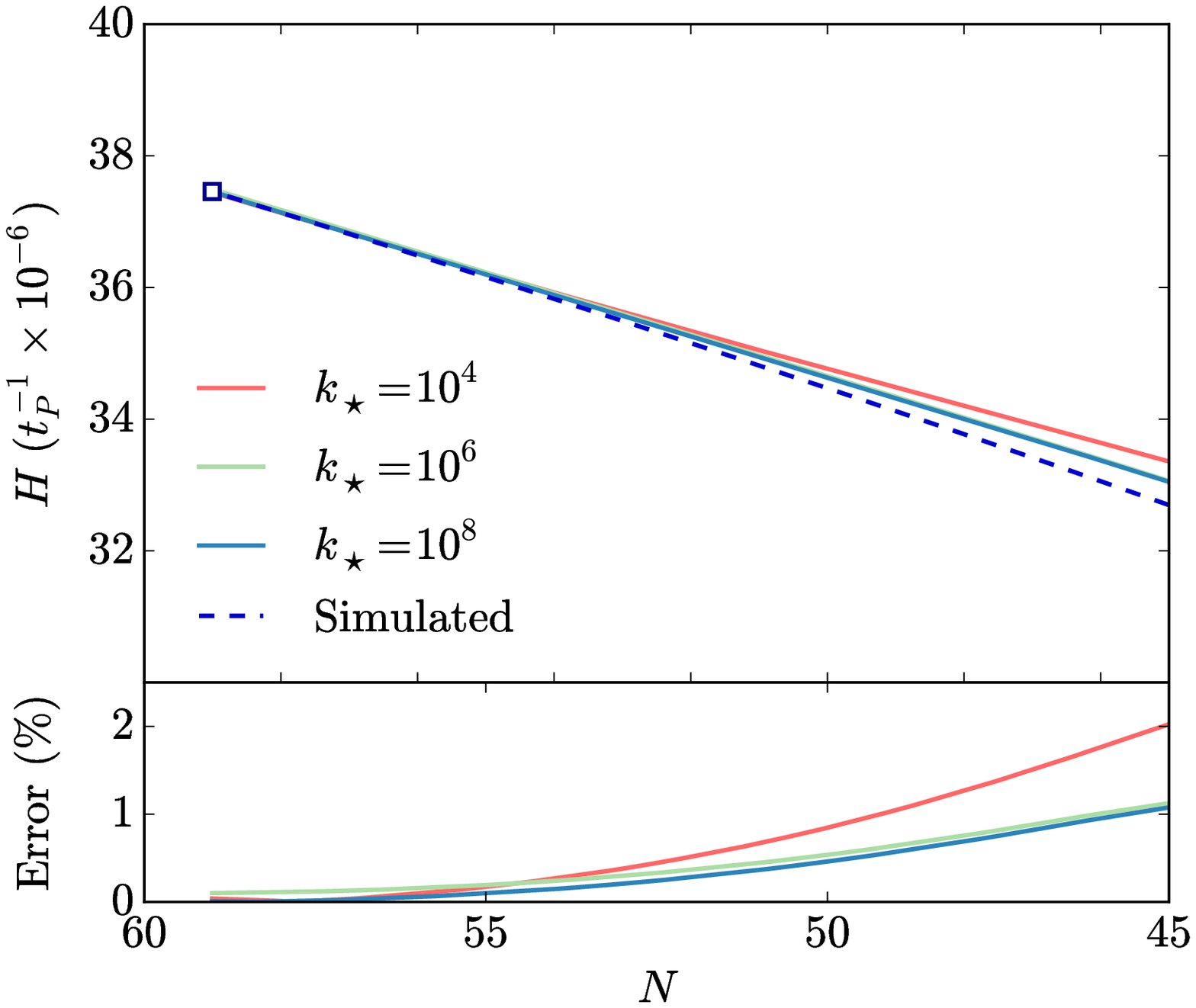}
  }
  \hspace*{\fill}
  \subfigure[\,Potential with a step]{
    \label{fig:Reconstructed_Hubble_Parameter_Step}
    \includegraphics[scale=0.45]{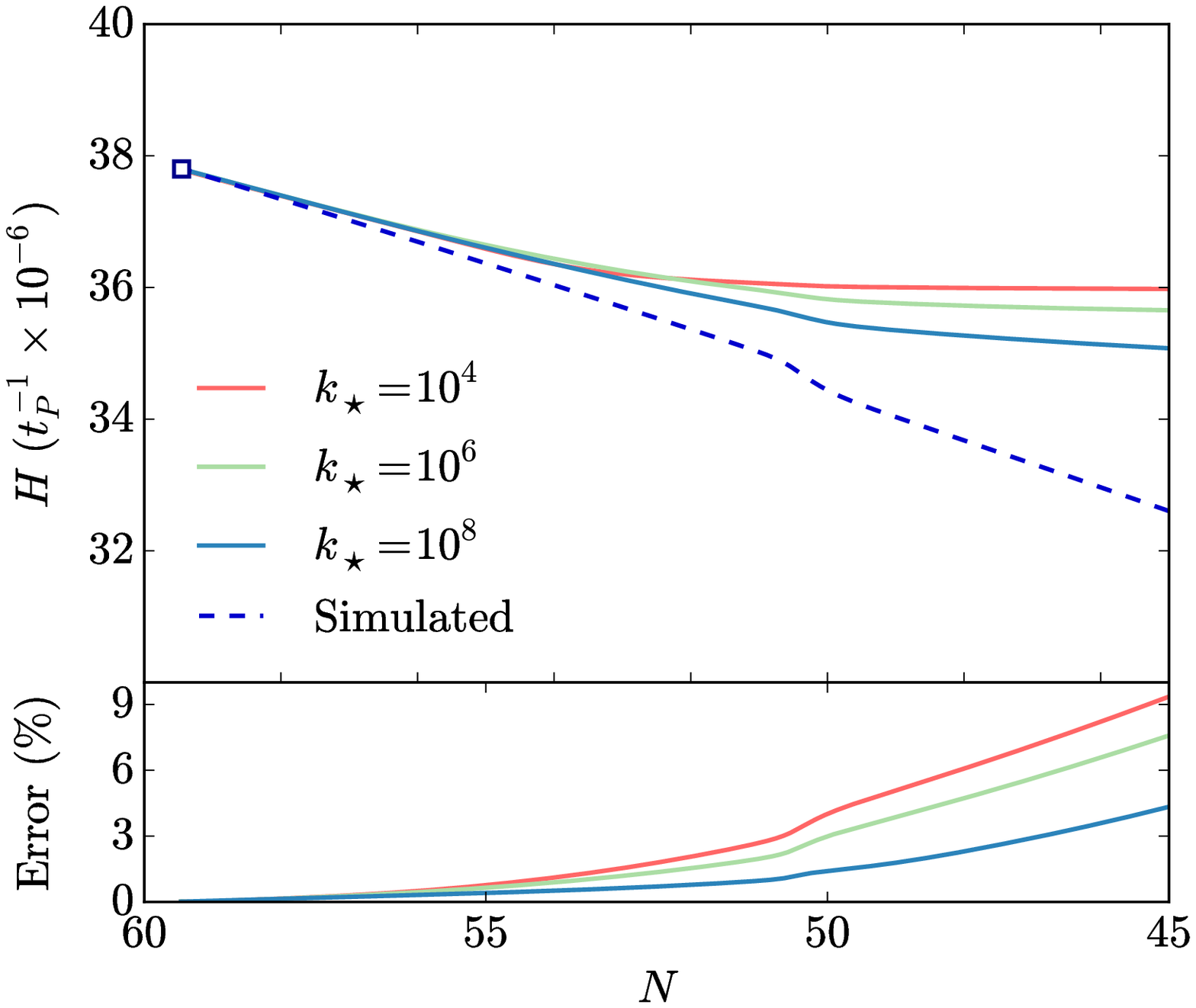}
  }
  \caption{Hubble parameter associated with the potential $V_{A}(\varphi)$ (left panels) and  $V_{B}(\varphi)$ (right panels) for different choices of the cut-off scale, $k_\star$ (in units of $\mathrm{Mpc^{-1}}$). For both cases, the upper panels shows the simulated (dashed lines) and the reconstructed (solid lines) curves for the Hubble parameter, $H$ (in units of the inverse Planck time), as a function of the number of e-folds, $N$. The relative error between the curves is shown in the lower panels. These reconstructions assume knowledge of the amplitude of $P_{T}(k)$ at $k_0 = 0.05\, \mathrm{Mpc^{-1}}$ (represented on the plots by a square marker at $N \approx 60$).
  }
  \label{fig:Reconstructed_Hubble_Parameter}
\end{figure*}

Figure \ref{fig:Reconstructed_Hubble_Parameter} shows both the reconstructed and simulated Hubble parameters for our choices of inflationary models expressed as a function of the number of e-folds, $N$. As expected, the reconstructed Hubble parameter is significantly more accurate close to $N \approx 60$, as the procedure relied on initial conditions extracted at CMB scales. The agreement between expected and reconstructed curves is markedly better for the model with a featureless quadratic potential and a choice of large cut-off scale. As for the potential with a step, the errors in the reconstructed $q(\eta)$ significantly affect the computation of $a(\eta)$, and consequently of $H$. In this case the reconstruction could benefit from improved $q(\eta)$ curves computed with higher numerical resolution, although the errors associated to the region containing the feature would still be large enough to affect the reconstruction for $N < 50$. Nevertheless, even though in this case the reconstructed curve may deviate significantly from the simulated solution, it is still possible to identify in Fig.~\ref{fig:Reconstructed_Hubble_Parameter_Step} the presence of a step in the evolution of $H$ at $N \approx 50$ for $k_\star = 10^{8} \,\mathrm{Mpc^{-1}}$.

\section{Future Prospects}\label{sec:Conclusions}

A formal analogy between determining inflationary dynamics from metric perturbations and determining a spherically symmetric quantum scattering potential from scattering data has been known for some time. We have obtained a numerical reconstruction of inflation dynamics given a known scalar perturbation power spectrum via the Gel'fand-Levitan method: First, take a primordial scalar power spectrum as input data and from it compute the Jost function. Second, numerically evaluate a highly oscillatory intergral to obtain the input kernel. Third, convert the Gel'fand-Levitan-Marchenko equation into a system of linear equations to obtain the freeze-out horizon as a function of scale from the derivative of the output kernel. Fourth, solve a set of differential equations to obtain the evolution of the Hubble parameter during inflation from the freeze-out horizon assuming we know the amplitude of the primordial tensor power spectrum. We performed these calculations for both a standard slow-roll inflation model and a model with an inflaton potential step resulting in a period of fast-roll dynamics, obtaining reasonably good agreement between simulated and reconstructed freeze-out horizon (Fig.~\ref{fig:Reconstructed_changing_freezeout}) and Hubble parameter (Fig.~\ref{fig:Reconstructed_Hubble_Parameter}). We have thus demonstrated that numerical computation of arbitrary inflation dynamics using the Gel'fand-Levitan method  can be numerically stable and accurate over the wide range of scales for which inflation observables can potentially be observed.  We use the assumption that the potential was slow-rolling in the infinite past in order to establish the value of $\ell$ when solving for the freeze-out horizon.

This calculation can form the basic building block for model-independent estimation of inflation dynamics from noisy measurements of the primordial density power spectrum. For a given set of power spectrum measurements and covariances, the most likely inflation dynamics is a well-posed Bayesian inference problem, since a calculation like the one here maps any given power spectrum into a particular inflation history. An additional possible future source of information is the power spectrum of tensor perturbations produced during inflation; the amplitude of a tensor spectrum depends on the energy scale at which inflation occurred. If large enough to be detected at one or more length scales, the tensor perturbations will provide additional constraints which can be folded into the model inference or incorporated as priors on the inflation effective potential obtained from the scalar perturbations. For instance, measurement of tensor modes at both CMB and Solar System scales by a space-based interferometer \cite{Caligiuri:2015,Turner:1996ck} could be used as boundary conditions for the inference of inflationary history at intermediate scales. Future surveys may measure the scalar power spectrum over a range up to $k \sim 10^6 \, {\rm Mpc}^{-1}$ \cite{Chluba:2012we, Keeton:2009, Ngan:2014}, however, the initial conditions (the amplitude of tensor perturbations) for solving the boundary value problem will be larger than observational range of the scalar power spectrum and this will require an statistical inference outside the limits of future experiments. Inflation dynamics estimation from noisy and incomplete power spectrum data will be considered elsewhere.

The basic technique of inverting observations to get estimates of the inflation dynamics is opposite to the forward modeling which is generally used to constrain inflation: given an inflation model, its predicted perturbation power spectra can be computed in a straightforward manner, and then compared to data. Many inflation models can be ruled out this way. But it is difficult to quantify which inflation model is most likely, because any parameterization of the space of models has no natural measure for prior probability.

The wider the range of length scales covered by power spectrum measurements, the longer the period of inflationary dynamics that can be recovered.  Current cosmic microwave background experiments, large-scale structure surveys, and Lyman-alpha observations give a  reasonably precise measurement of the primordial power spectrum for wave numbers between  $0.01\,\mathrm{Mpc^{-1}} <  k < 1 \, \mathrm{Mpc^{-1}}$ \cite{Sealfon, Verde, Guo, Hu, Hlozek, Vazquez, Planck_Inflation}.  Microwave background spectrum distortions in principle contain information about perturbations at smaller scales \cite{Chluba:2012we}. It is also possible that primordial perturbations on subgalactic scales will eventually be probed by high-sensitivity gravitational lensing, galaxy dynamics measurements or improved understanding of the dwarf galaxy population. Tensor perturbations may be seen at horizon scales in microwave background B-mode polarization, and if the tensor amplitude is large enough to be detected this way, then ultimately the tensor perturbations may also be measured at a vastly smaller Earth-Sun scale by a space-based laser interferometer \cite{Turner:1996ck, Caligiuri}. 

Some of these signals may be beyond the ultimate reach of experiments, but long-term prospects exist for significantly enhanced experimental information about inflation-produced perturbations. Inverting the data to obtain constraints on inflation dynamics will allow us to make the most of these remarkable observational possibilities. 

\section{Acknowledgments}
F.Z. thanks Gerard Jungman and Salman Habib for helpful discussions. J.M. gratefully acknowledges CONACyT for supporting this work. The authors have been partly supported by NSF grant AST-1312380. This work made use of the GNU Scientific Library \cite{galassi1gnu}, the NumPy \cite{van2011numpy}, SciPy \cite{jones2001open}, and Matplotlib \cite{hunter2007matplotlib} Python libraries, as well as the PyGSL \cite{gaedke2005pygsl} Python wrapper. Bibliographic information was obtained from the NASA Astrophysical Data System.

\appendix

\section{The Jost Function}\label{app:Jost_Function}
The Jost function, $F_{\ell}(k)$, encapsulates all the information about the scattering problem and therefore plays a central role in the theory of inverse scattering. Its definition is given in terms of two particular sets of solutions for the the radial time-independent Schr\"{o}dinger equation: the regular solutions, $\varphi_{\ell}(k, r)$; and the Jost solutions, $f_{\ell}(k, r)$. Regular solutions are those satisfying the boundary condition $\lim_{r \rightarrow 0} (2\ell + 1)!!r^{-\ell - 1}\varphi_{\ell} = 1$. The Jost solutions, on the other hand, are those satisfying  $\lim_{r \rightarrow \infty}e^{-ikr}f_{\ell}(k, r) = 1$. This two regular solutions may joint in the Jost function defined by the Wronskian
\begin{equation*}
F_\ell(k) = (-k)^{\ell}W\{f_{\ell}(k, r), \, \varphi_{\ell}(k, r)\} \, .
\end{equation*}
Taking the limit of this expression for $r \rightarrow 0$, we obtain \cite{Newton_Chadan_Sabatier}
\begin{equation}\label{eq:Appendix_Jost_Function}
F_{\ell}(k) = \lim_{r \to 0}\left[ \frac{e^{-i\pi\ell}\Gamma\left( \frac{1}{2} \right)}{\Gamma\left( \ell + \frac{1}{2} \right)}\left( \frac{k r}{2} \right)^{\ell}  f_\ell(k,r) \right] \, .
\end{equation}

In the context of inflationary perturbations, the behavior of the Jost solutions  as $\eta \rightarrow -\infty$ coincides with that of modes set by the Bunch-Davies vacuum. As a consequence, the Jost solutions must behave as $f_{\ell}(k, r) = A_k z$ for $r \rightarrow 0$. Furthermore, for small $r$, the effective potential experienced by the scattered particle is mostly due to the centrifugal term $\ell(\ell + 1)r^{-2}$ in the Schr\"{o}dinger equation (\ref{eq:Schrodinger_Scattering_Equation}). Therefore, for $r \rightarrow 0$ the Mukhanov variable must have the form  $z = r_{0}/r^{\ell}$, where $r_0$ is a constant. Combining these results with (\ref{eq:Appendix_Jost_Function}) gives the following expression relating the Jost function and $A_k$:
\begin{equation*}
	A_k = \frac{\Gamma(\ell+\frac{1}{2})}{\pi r_0}\left( \frac{2}{k} \right)^{\ell+\frac{1}{2}} F_{\ell}(k) \, .
\end{equation*}
This expression, along with Eq.~(\ref{eq:Primordial_PS}), yields Eq.~(\ref{eq:Jost_Function}) which relates the Jost function to the primordial power spectrum.

\section{Solving the Gel'fand-Levitan-Marchenko Equation}\label{app:Solving_GLM}
Below we describe the method of separable kernel decomposition for solving the Gel'fand-Levitan-Marchenko equation. We start by assuming that the input kernel, $G(r,s)$, can be written as
\begin{equation}\label{eq:Separable_Kernel}
	G(r,s) = \sum_{i=1}^{m} a_{i}(r) b_{i}(s) \, ,
\end{equation}
where $a_i = \mathit{j}_{\ell}(rk_{i})$ and $b_{i} = \mathit{j}_{\ell}(sk_{i}) \left[ |F_{\ell}(k_{i})|^{-2}-1 \right] \Delta_i$. Here, $\Delta_i$ essentially represents the integration step. For higher numerical resolution $\Delta_{i} \rightarrow 0$, and consequently $m \rightarrow \infty$. Now consider the Gel'fand-Levitan-Marchenko equation:
\begin{equation*}\label{eq:Appendix_GLM_Equation}
	K(r,s) + G(r,s) + \int_0^r K(r,t) G(t,s) dt = 0 \, .
\end{equation*}
We can make use of Eq.~(\ref{eq:Separable_Kernel}) to rewrite it as
\begin{eqnarray}
	K(r,s) &=& - G(r,s) - \sum_{i=1}^{m} b_i(s) \int_0^r K(r,t) a_i(t) dt \label{eq:Appendix_GLM_Equation_2} \\
	K(r,s) &=& -G(r,s) - \sum_{i=1}^{m} b_{i}(s)c_{i}(r) \, \label{eq:Appendix_GLM_Equation_3},
\end{eqnarray}
where we have defined
\begin{equation}\label{eq:Definition_c}
  c_{i}(r) \equiv \int_0^r K(r,t)a_{i}(t) dt \, .
\end{equation}
Replacing this last expression for $K(r,s)$ in (\ref{eq:Definition_c}) we obtain a system of $m$ equations given by
\begin{equation}\label{eq:Appendix_GLM_Linear_System}
	c_{i}(r) = -g_{i}(r) - \sum_{j=1}^{m} c_{j}(r)h_{ij}(r) \, ,
\end{equation}
where we have defined
\begin{equation*}
\begin{split}
	g_{i}(r) &\equiv \int_0^r G(r,t) a_{i}(t) dt \\
	h_{ij}(r) &\equiv \int_0^r a_{i}(t) b_{j}(t) dt \, .
\end{split}
\end{equation*}

Since $G(r,s)$ is known and both $a_{i}$ and $b_{i}$ can be easily computed, the Gel'fand-Levitan-Marchenko equation is reduced to the linear system of equations (\ref{eq:Appendix_GLM_Linear_System}). The solutions for $c_i(r)$ can easily be obtained numerically and corresponds to an $m$-tuple: $(c_1,...,c_m)$. Substituting this result in Eq.~(\ref{eq:Appendix_GLM_Equation_3}) gives the desired output kernel $K(r,s)$.

\section{The Levin Method}\label{app:Levin_Method}
Here we show the essential results needed for an implementation of the Levin method targeted at computing the oscillatory integral in Eq.~(\ref{eq:Input_Kernel}) for the input kernel. For a more detailed explanation of this integration method we refer the reader to the original paper by Levin \cite{Levin}. Below we follow Levin's notation.

First, we assume that the integral Eq.~(\ref{eq:Input_Kernel}) has a solution of the form
\begin{equation}
    \int \mathbf{f}(k) \cdot \mathbf{w}(k) dk = \sum_i^m F_i(k) w_i(k),
\end{equation}
where $\mathbf{w}(x)$ is defined to be the vector of length $m$ containing the oscillatory kernel function. In our case it is clearly
\begin{equation*}\label{eq:Levin_Vector_w}
    \mathbf{w}(k) = \begin{bmatrix}
           J_{\ell}(r k)J_{\ell}(s k) \\
           J_{\ell - 1}(r k)J_{\ell}(s k) \\
           J_{\ell}(r k)J_{\ell - 1}(s k) \\
           J_{\ell - 1}(r k)J_{\ell - 1}(s k)
         \end{bmatrix} \, ,
\end{equation*}
with the amplitude function 
\begin{equation*}\label{eq:Levin_Vector_f}
	\mathbf{f}(k) = \begin{bmatrix}
           k \left[ |F_{\ell}(k)|^{-2} - 1 \right] \\
           0 \\
           0 \\
           0
         \end{bmatrix} \, .
\end{equation*}
We use a Chebyshev polynomial expansion for the solution function, $F_i(x) = \sum_k^n c_{ik} u_k(x)$. The final step is to compute the ($m \times n$) $c_{ik}$ coefficients by solving the following ordinary differential equation
\begin{equation}
    f_i(x) = F_i^\prime(x) + \sum_j^m A_{ij}(x)F_j(x) \, ,
\end{equation} 
where the matrix $A_{ij}$ must satisfy the relation $d\mathbf{w}(k)/dk = \mathbf{A}(k)\mathbf{w}(k)$. Using the Bessel function recurrence relations, the matrix $\mathbf{A}$ is given by
\begin{equation*}\label{eq:Levin_A_matrix}
	\mathbf{A} = \begin{bmatrix}
           -\frac{2\ell}{k}     & r            & s            & 0 \\
           -r                   & -\frac{1}{k} & 0            & s \\
           -s                   & 0            & -\frac{1}{k} & r \\
           0                    & -s           & -r           & \frac{2(\ell - 1)}{k}
         \end{bmatrix} \, .
\end{equation*}

Employing these expressions for the Levin's collocation procedure described in \cite{Levin} allows one to compute the integral in Eq. (\ref{eq:Input_Kernel}). The convergence of this integration method depends roughly on the ratio between the size of the integration range and the $n$ number of collocation points, with better convergence being achieved for smaller values of this ratio.

The accuracy of the inversion procedure is directly affected by step size in the numerical evaluation of the output kernel. More accurate results can be obtained by increasing the numerical resolution in the inversion algorithm, at the cost of increasing the computation time. In our calculations we employ a resolution of $10^{4}$ bins for the smallest logarithmic region, $[10^{-6},\, 10^{-5}] \, \mathrm{Mpc}^{-1}$, when computing the integral Eq.~(\ref{eq:Input_Kernel}) and keep the resolution constant for the rest of the analysis. A more detailed error analysis of the reconstruction will be investigated in a future work.


\bibliographystyle{unsrtnat}
\bibliography{Inverse_Scattering_Paper}

\end{document}